\documentstyle[aps,manuscript,amsfonts]{revtex}
\begin{document}
\title{Magnetoelastic coupling in epitaxial
magnetic films: An ab-initio study}
\author{M. Komelj\\{\it\footnotesize Jozef Stefan Institute, Jamova 39, 1000
Ljubljana, Slovenia}\\
M. F\" ahnle\\
{\it\footnotesize Max Planck-Insitut f\" ur Metallforschung, Heisenbergstra\ss e
1, D-70569 Stuttgart, Germany}}
\maketitle
\begin{abstract}
A method is developed which allows to determine the first-order and
the second-order magnetoelastic coefficients of a magnetic bulk material 
from the 
ab-initio calculation of the magnetocrystalline anisotropy energy as
function of a prestrain $\epsilon_0$. Explicit results are given for bcc Fe,
and they agree well with experimental data obtained from the magnetostrictive
stress measurements for epitaxial Fe films. 
\vspace{12pt}
\end{abstract}
\newpage
In recent years magnetic devices based on magnetic films technologies 
have attracted a considerable interest, e.g., magnetooptical recording
media or magnetoresistive devices based on the giant magnetoresistive and
the tunnel magnetoresistive effect designed for sensors or magnetostrictive
random access memories. Thereby the magnetic anisotropy plays an important
role, for instance, the issue of perpendicular anisotropy for the 
magnetooptical recording or the demand for soft magnetic layers with weak
anisotropy as part of the magnetoresistive devices. It has been 
shown by numerous investigations that the magnetic anisotropies 
of magnetic films grown epitaxially on a substrate may strongly deviate
from those of the respective bulk materials. The reason for this deviation 
is in general ascribed to several different effects. First, there are 
contributions to the anisotropy originating from the free surface 
of the magnetic layer and from the interface between layer and substrate, 
as well as from the morphology of the film due to a heterogeneous film
growth. The influence of all these effects must decrease with increasing
thickness of the film. What remains for a film of thickness  larger than 
typically $10\>{\rm nm}$ is the effect
of the magnetoelastic coupling to the film strain induced by the lattice
mismatch between film and substrate. Because the epitaxial film strain may
be of the order of several \% which is much larger than typical magnetostrictive
strains of $10^{-6}$ to $10^{-4}$ and because of the dependence of
the magnetoelastic coupling energy on the direction of the magnetization
(see below),
this may result in new magnetic anisotropies different from that of the 
unstrained bulk. The numerous experiments on the effect of epitaxial strain
on the magnetic properties of magnetic films are reviewed in 
Ref. \cite{1}. An
ab-initio study of this effect within the framework of density functional 
theory is the purpose of the present letter. \par
To be more specific we 
consider a material which is cubic in the unstrained state (Fe, for instance).
Then the density of the magnetoelastic coupling energy may be written (up 
to the second order in the strain $\epsilon_{ij}$, omitting
the terms including the shear strain $\epsilon_{ij},i\ne j$, which are
not required for the situation discussed below) as \cite{2}
\begin{eqnarray}
&f_{\rm me}=B_1\left(\epsilon_{11}\alpha_1^2+\epsilon_{22}\alpha_2^2+
\epsilon_{33}\alpha_3^2\right)+\\
\nonumber
&{1\over 2}D_{11}\left(\epsilon_{11}^2\alpha_1^4+\epsilon_{22}^2\alpha_2^4+
\epsilon_{33}^2\alpha_3^4\right)+D_{12}\left(\epsilon_{11}\epsilon_{22}
\alpha_1^2\alpha_2^2+\epsilon_{22}\epsilon_{33}\alpha_2^2\alpha_3^3+
\epsilon_{33}\epsilon_{11}\alpha_3^2\alpha_1^2\right).
\end{eqnarray}
Here $B_1$ and $D_{11}$, $D_{12}$ represent magnetoelastic 
coupling coefficients of the first and the second order, and the $\alpha_i$
denote the direction cosines of the magnetization referred to the cubic axes. 
For Fe grown epitaxially on a cubic (100) surface which is a prototype 
system experimentally investigated intensively \cite{1,3,4} there 
are epitaxial strains $\epsilon_{11}=\epsilon_{22}$ which may be different 
for various atomic layers of the film \cite{1,4}. In the following we adopt 
a simple model where we consider only the average strain of the film which
depends on the thickness of the film, i.e., we assume 
$\epsilon_{11}=\epsilon_{22}=\epsilon_0$. The strain $\epsilon_{33}$ 
then may be obtained by minimizing the total energy density $f=f_{\rm me}+
f_{\rm el}$, with the elastic energy density (again omitting the 
terms containing the shear strains):
\begin{equation}
f_{\rm el}={1\over 2}C_{11}\left(\epsilon_{11}^2+\epsilon_{22}^2+\epsilon_{33}^2
\right)+
C_{12}\left(\epsilon_{11}\epsilon_{22}+\epsilon_{22}\epsilon_{33}+
\epsilon_{33}\epsilon_{11}\right)
\end{equation}
and with the cubic elastic stiffness constants $C_{ij}$. This yields
\begin{equation}
\epsilon_{33}=-{2C_{12}\epsilon_0+B_1\alpha_3^2\over C_{11}+D_{11}
\alpha_3^4}.
\end{equation}
Because the magnetostrictive contribution to the strain originating 
from $f_{\rm me}$ is much smaller than $\epsilon_0$ we find $\epsilon_{33}
\approx -2C_{12}/C_{11}$. \par
From (1), (3) we obtain the strain dependent part, $f_{\rm mca}(\epsilon_0)$,
of the magnetocrystalline anisotropy energy density:
\begin{equation}
f_{\rm mca}(\epsilon_0)=f_{\rm me}(\epsilon_0,\alpha_1=1)-f_{\rm me}(\epsilon_0,
\alpha_3=1)=k_0+k_1\epsilon_0+k_2\epsilon_0^2, 
\end{equation}
with
\begin{eqnarray}
k_0=&{1\over 2}B_1^2{1\over C_{11}+D_{11}},\\
k_1=&B_1\left(1+2C_{12}{1\over C_{11}+D_{11}}\right),\\
k_2=&{1\over 2}D_{11}\left(1-4{C_{12}^2\over C_{11}}{1\over C_{11}+D_{11}}
\right).
\end{eqnarray}
The term $k_0$ is negligible and arises from the fact that we fix
the strains $\epsilon_{11}=\epsilon_{22}=\epsilon_0$ independently on 
the direction of the magnetization. Furthermore, when changing the direction
of the magnetization there is a change in the magnetostrictive stress
$\tau_1={\partial f_{\rm me}\over\partial\epsilon_{11}}$ according 
to 
\begin{equation}
\Delta\tau_1=\tau_1(\alpha_1=1)-\tau_1(\alpha_2=1)=B_1+D_{11}\epsilon_0.
\end{equation}
An experimental determination of $\Delta\tau_1$ (exploiting the
change of the bending moment that is created by the film onto
the substrate \cite{1,3,4}) as function of the layer thickness
and hence as function of $\epsilon_0$ then yields the two magnetoelastic
coupling coefficients $B_1$ and $D_{11}$. For Fe on MgO (100) 
Koch {\it et al.} \cite{3} obtained $B_1=-3.2\>{\rm MJ/m^3}$, 
$D_{11}=1.1\>{\rm GJ/m^3}$ $(\pm10\%)$,
and for Fe on W(100) Enders {\it et al.} \cite{4} found $B_1=-3\>{\rm MJ/m^3}$, 
$D_{11}=1\>{\rm GJ/m^3}$. 
The values for $B_1$ extracted from the film experiments agree rather
well with the bulk value of $B_1=-3.44\> {\rm MJ/m^3}$. There are no values of 
$D_{11}$
obtained  from bulk measurements for comparison and therefore the two 
experiments were considered as the first determination of the second-order 
magnetoelastic coupling coefficient of bulk Fe by a film experiment. \par
It was pointed out \cite{1,3,4} that due to the large strains accessible 
by epitaxial film growth the effective first-order coefficient
$B_{\rm eff}$ defined as $B_{\rm eff}=B_1+D_{11}\epsilon_0$ changes
sign from negative to positive for Fe with decreasing film thickness, i.e.,
increasing $\epsilon_0$, and this clearly demonstrates that the magnetic 
anisotropy energy depends dramatically on the film thickness, a result
which is most relevant for the design of the magnetic film devices (see
introduction). The linear strain dependence of $B_{\rm eff}$ failed to 
describe the experimental data for film thicknesses below $10\>{\rm nm}$, most 
probably because 
then the effects of the surface, interface and film morphology 
become relevant. \par
In the present paper we determine the magnetoelastic coupling coefficients
$B_1$ and $D_{11}$ for cubic Fe by the ab-initio density functional theory. 
To do this, we calculate the magnetocrystalline anisotropy energy density
$f_{\rm mca}(\epsilon_0)$ as function of the strain $\epsilon_{11}=\epsilon_{22}
=\epsilon_0$ imposed to the bulk material, represent the data by a quadratic polynomial in $\epsilon_0$
according to eq. (4) and determine $B_1$ and $D_{11}$ from eqs. (6,7), 
inserting the elastic stiffness constants $C_{12}$ and $C_{11}$ which 
we have also obtained ab initio.\par 
We have performed the calculations using the WIEN97 \cite{5}
code which adopts the full-potential linearized augmented plane-wave
(FLAPW) method \cite{6}. For the exchange-correlation potential  
the local-spin-density (LSDA) functional by Perdew and Wang \cite{7} and the 
generalized-gradient-approximation (GGA) functional by Perdew {\it et al.} 
\cite{8} 
were used. 
The total energy minimizations on the non-strained bcc Fe gave us
the equilibrium lattice parameters $a=5.2\>{\rm a_0}$ for LSDA and 
$a=5.34\>{\rm a_0}$ for 
GGA where ${\rm a_0}$ denotes Bohr's radius. The calculated ratio $2C_{12}/C_{11}$
is 1.08 for LSDA and 1.13 for GGA. The experimental values are 
$a=5.42\>{\rm a_0}$ and $2C_{12}/C_{11}=1.17$. \par
Numerically the most difficult 
step is the calculation of $f_{\rm mca}$ which is due to the spin-orbit 
coupling 
(SOC).  First, we calculate the self-consistent electronic structure in
the scalar-relativistic approximation \cite{9} using $N_{\bf k}^3$ 
${\bf k}$ vectors with $N_{\bf k}=21$ in the total Brillouin 
zone (BZ) which correspond to the 762 ${\bf k}$ vectors in the irreducible
part of the Brillouin zone (IBZ). 
The criterion for the self-consistency is the difference 
in the charge densities after the last two iterations being less 
than $2\times 10^{-6}e/({\rm a.u.})^3$. The contribution of the SOC
is determined perturbatively  using the second variational method \cite{10,11}.
The quantity $f_{\rm mca}$ is calculated by applying the force theorem 
\cite{12,13} as the 
difference between the sums of the perturbed eigenvalues for the different 
magnetization directions. Fig. 1 represents the convergency test for the
calculation of $f_{\rm mca}$ with respect to $N_{\bf k}$. The data are for the 
case
with $a=5.4\>{\rm a_0}$ and $c=5.2\>{\rm a_0}$, using GGA. The modified 
tetrahedron \cite{14}
and the Gaussian smearing \cite{15,16} integration schemes were used. 
The proper convergency with the Gaussian smearing was achieved by setting the 
smearing parameter as $\Gamma/N_{\bf k}$. 
The suitable values of $\Gamma$ for the particular case are roughly from
the interval between $6.8\>{\rm eV}$ and $10.1\>{\rm eV}$ which follows from 
the curves in 
Fig. 1. All the final calculations of $f_{\rm mca}$ were performed with 
$N_{\bf k}=51$ (17576 ${\bf k}$ 
vectors in the IBZ) using both the tetrahedron and the Gaussian smearing 
method in order to minimize the numerical uncertainties. The estimated accuracy 
is $\pm 1\>{\rm \mu eV/unit\, cell}$ marked by the horizontal lines in Fig. 1.
\par
Fig. 2 shows the calculated magnetocrystalline anisotropy energy 
density $f_{\rm mca}$ 
with respect to the lateral strain $\epsilon_0$. The numerical data are 
well fitted by quadratic polynomials as predicted by eq. (4). From the calculated 
parameters $k_1$ 
and $k_2$ the magnetoelastic coefficients $B_1$ and $D_{11}$ are determined
according to eqs. (6,7). 
Table I summarizes the theoretical results in comparison with the experimental
data from \cite{1,4}.  
There is a big discrepancy between the LSDA result for $B_1$ and the 
experimental 
result, whereas the GGA result is much closer to the experiment. This 
is in line with the calculation of the magnetoelastic coefficient 
$\lambda_{100}$ of unstrained bulk Fe by Wu {\it et al.} \cite{17} who also 
obtained a strong deviation from the experiment when using LSDA but a 
satisfactory agreement when using GGA. 
The calculated
second-order magnetoelastic coupling coefficient $D_{11}$ for bulk bcc Fe 
matches the experimental value obtained from the measurements on epitaxial
thin films very well, especially the value from the GGA calculation. 
The agreement represents the direct proof that the
experimental results \cite{1,3,4} can be really ascribed to 
the pure strain effect on the magnetoelastic properties and that the
measurements of the magnetostrictive film stress as function of the film
thickness can provide the second-order coupling constant $D_{11}$ of
the bulk which is hard to obtain by bulk measurements. \par
We close with an important warning. The experiments \cite{1,3,4} and
the present theory demonstrate that the magnetoelastic properties of
thin epitaxial films may deviate significantly from that described by 
the first-order magnetoelastic coupling coefficients of the bulk and that 
one has to take into account the second-order terms of the bulk. The 
change of the magnetostrictive stress $\Delta\tau_1$ obtained when
switching the magnetization from [010] to [100] as function of the 
epitaxial strain $\epsilon_0$ then may be expressed by an effective
first-order coefficient $B_{\rm eff}=B_1+D_{11}\epsilon_0$. However,
the correct result for the coefficients $k_0$, $k_1$ and $k_2$ of the 
polynomial expansion (4) of $f_{\rm mca}$ may {\it not} be obtained by 
neglecting the second-order terms in eq. (1) and instead replacing  $B_1$ in 
the first-order term by $B_{\rm eff}=B_1+D_{11}\epsilon_0$. 
This would yield $f_{\rm mca}=k_0+k_1\epsilon_0+k_2\epsilon_0^2$ with
$k_1=B_1(1+2C_{12}/C_{11})$ and $k_2=D(1+2C_{12}/C_{11})$ instead
of eqs. (6,7). Inserting the value of $2C_{12}/C_{11}$ for Fe it 
becomes obvious that $D$ and $D_{11}$ even have a different sign;
the sign of $D$ being opposite to the one of the experimentally determined
$D_{11}$ according to eq. (8). Guo {\it et al.} \cite{18} have performed 
an ab-initio calculation of the magnetoelastic properties of 
epitaxial Co and Ni films, and they indeed proceeded on this line, i.e.
they neglected the second-order term in eq. (1) and instead 
replaced the constant $B_1$ by a strain-dependent term $B_1(\epsilon_0)$. 
It should be cautioned that the coefficient $D$ obtained from the 
linearization of $B_1(\epsilon_0)$ is not identical to the second 
order coefficient $D_{11}$ of the bulk material but it may deviate 
strongly.\par
\vspace{12pt}
{\bf Acknowledgement}: The authors are indebted to O. Grotheer, P. Novak and
R.Q. Wu for helpful discussions.

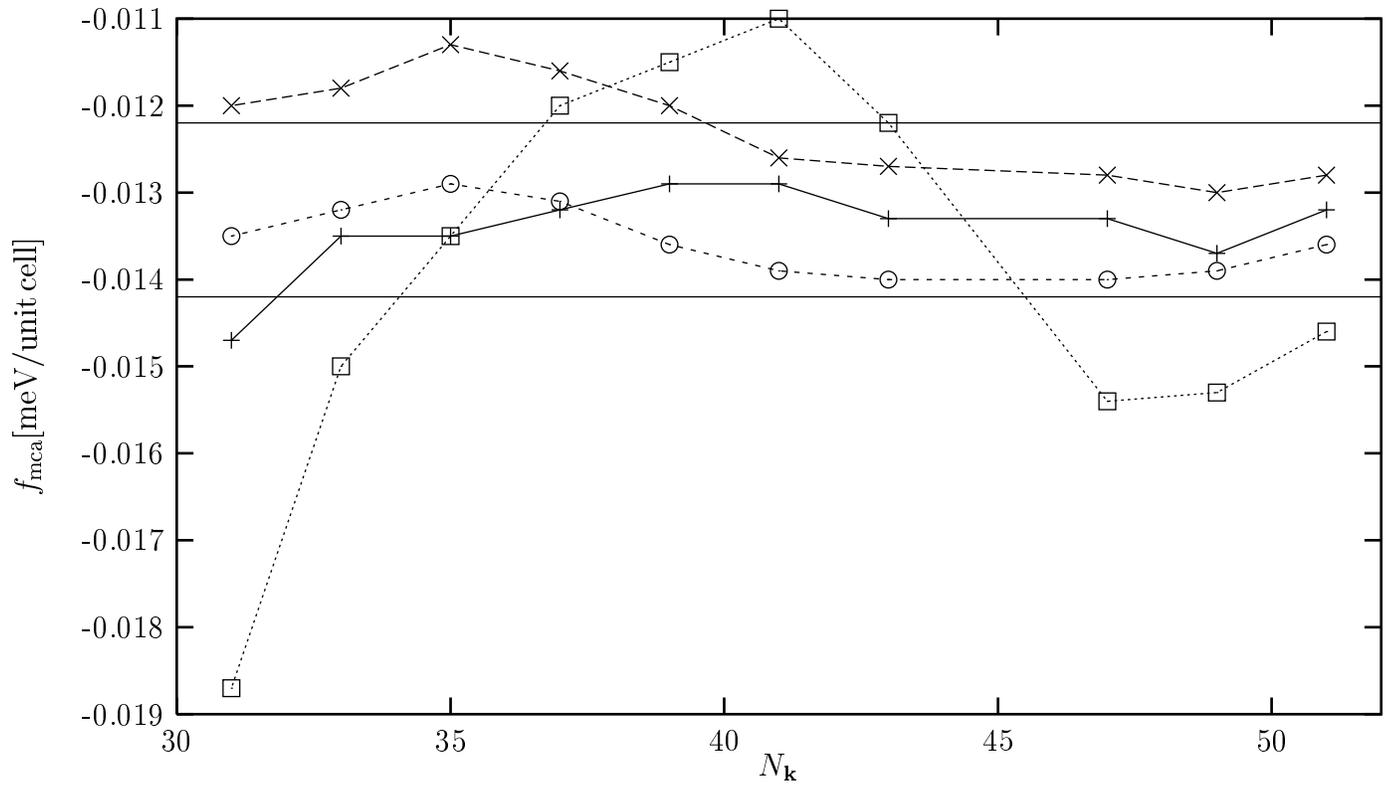
\begin{figure}
\caption{The convergency test for $f_{\rm mca}$ with respect to 
$N_{\bf k}$.
The horizontal lines represent the estimated accuracy 
$\pm 1\,{\rm \mu eV}/{\rm unit\,cell}$; $+$, modified tetrahedron method; $\times$, $\circ$ and $\square$,
Gaussian smearing with $\Gamma=10.2\>{\rm eV}$, $6.8\>{\rm eV}$ and $1.02\>
{\rm eV}$.
The proper convergency using the Gaussian smearing is obtained 
for the smearing parameter $\Gamma\over N_{\bf k}$ where $\Gamma$ is
from the interval between $6.8\>{\rm eV}$ and $10.1\>{\rm eV}$.}
\end{figure}
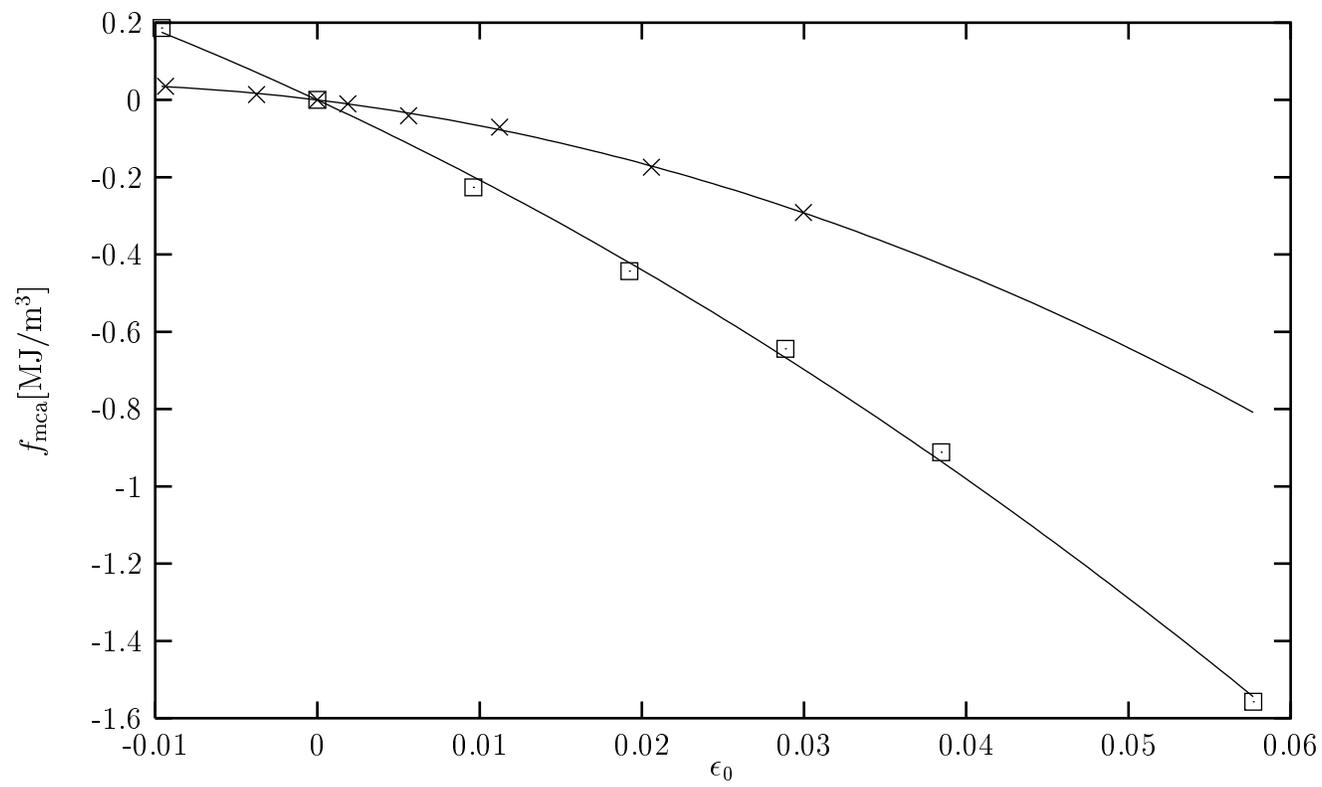
\begin{figure}
\caption{The calculated magnetocrystalline anisotropy energy density 
$f_{\rm mca}$ as function of the prestrain $\epsilon_0$.
The solid lines represent fits to quadratic polynomials 
according to eq. (4). Note that $k_0$ is indeed negligible; $\square$, LSDA;
$\times$, GGA.}
\end{figure}
\begin{table}
\begin{center}
\caption{The calculated results in comparison with the experimental 
data from the thin film experiments.}
\begin{tabular}{lrrlc}
&$a{\rm [a_0]}$&$2{C_{12}\over C_{11}}$&$B_1{\rm [MJ/m^3]}$&$D_{11}{\rm [GJ/m^3]}$\\
\hline
LSDA&5.20&1.08&-9.26&1.5\\
GGA&5.34&1.13&-2.39&1.1\\
exp.\cite{1,4}&5.42&1.17&-3&1.0\\
exp.\cite{3}  &    &    &-3.2&1.1\\
\end{tabular}
\end{center}
\end{table}

\newpage
\begin{thebibliography}{99}
\bibitem{1} D. Sander, Rep. Prog. Phys. {\bf 62}, 809 (1999).
\bibitem{2} R.C. O'Handley and S.-W. Sun, J. Magn. Magn. Mat. {\bf 104-107} 1717 
(1992).
\bibitem{3} R. Koch, M. Weber K. Th\" urmer and  K.H. Rieder, 
J. Magn. Magn. Mat. {\bf 159}, L11 (1996);
G. Wedler, J. Walz, A. Greuer and  R. Koch, Phys. Rev. B {\bf 60}, R11313 
(1999).
\bibitem{4} A. Enders, D. Sander and  J. Kirschner, J. Appl. Phys. {\bf 85},
5279 (1999).
\bibitem{5} P. Blaha, K. Schwarz, P. Sorantin and  S.B. Trickey, 
Comput. Phys. Commun. {\bf 59}, 399 (1990).
\bibitem{6} E. Wimmer, H. Krakauer, M. Weinert and  A.J. Freeman, 
Phys. Rev. B {\bf 24}, 864 (1981).
\bibitem{7} J.P. Perdew and  Y. Wang, Phys. Rev. B {\bf 45}, 13244 (1992).
\bibitem{8} J.P. Perdew {\it et al.}, Phys. Rev. B {\bf 46},  6671 (1992).
\bibitem{9} D.D. Koelling and  B.N. Harmon, J. Phys. C: Solid State Phys. 
{\bf 10},
3107 (1977).
\bibitem{10} D. Singh, {\it Plane Waves, Pseudopotentials and the LAPW method}
(Kluwer Academic, Dordrecht 1994), p. 86.
\bibitem{11} P. Novak (unpublished).
\bibitem{12} A.R. Mackintosh and  O.K. Andersen, in {\it Electrons at the Fermi
Surface}, edited by M. Springford (Cambridge University Press, Cambridge 1980).
\bibitem{13} M. Weinert, R.E. Watson and J.W. Davenport, Phys. Rev. B {\bf 32} 
2115 (1985).
\bibitem{14} P.E. Bl\" ochl, O. Jepsen and  O.K. Andersen, Phys. Rev. B {\bf 49}
16223 (1994).
\bibitem{15} C.-L. Fu and  K.-M. Ho, Phys. Rev. B {\bf 28} 5480 (1983).
\bibitem{16} O. Grotheer and  M. F\" ahnle, Phys. Rev. B {\bf 58} 13459 (1998).
\bibitem{17} R.Q. Wu, L.J. Chen, A. Shick and  A.J. Freeman, J. Magn. Magn. Mat.
{\bf 177-181} 1216 (1998).
\bibitem{18} G.Y. Guo, J. Magn. Magn. Mat. {\bf 209} 33 (2000);
G.Y. Guo, D.J. Roberts and G.A. Gehring, Phys. Rev. B {\bf 59}, 14466 (1999).
\end{thebibliography}
\end{document}